\documentclass[%
 aip,
 amsmath,amssymb,
 reprint,%
]{revtex4-1}

\usepackage{graphicx}
\usepackage{dcolumn}
\usepackage{bm}

\usepackage[utf8]{inputenc}
\usepackage[T1]{fontenc}
\usepackage{mathptmx}
\usepackage{etoolbox}
\usepackage{comment}

\makeatletter
\def\@email#1#2{%
 \endgroup
 \patchcmd{\titleblock@produce}
  {\frontmatter@RRAPformat}
  {\frontmatter@RRAPformat{\produce@RRAP{*#1\href{mailto:#2}{#2}}}\frontmatter@RRAPformat}
  {}{}
}%
\makeatother
\begin{document}

\preprint{AIP/123-QED}

\title[Optimizing spintronics via gradient descent]{Gradient-based optimization of spintronic devices}
\author{Y. Imai}
\affiliation{ 
Graduate School of Information Science and Technology, The University of Tokyo, 7-3-1 Hongo, Bunkyo-ku, 113-8656 Tokyo, Japan
}%
\author{S. Liu}
\affiliation{ 
Graduate School of Information Science and Technology, The University of Tokyo, 7-3-1 Hongo, Bunkyo-ku, 113-8656 Tokyo, Japan
}%
\author{N. Akashi}%
 \affiliation{%
Graduate School of Informatics, Kyoto University, Yoshida-honmachi, Sakyo-ku, 606-8501 Kyoto, Japan
}%
\author{K. Nakajima}
\affiliation{ 
Graduate School of Information Science and Technology, The University of Tokyo, 7-3-1 Hongo, Bunkyo-ku, 113-8656 Tokyo, Japan
}%

\email{imai@isi.imi.i.u-tokyo.ac.jp}

\date{\today}

\begin{abstract}
The optimization of physical parameters serves various purposes, such as system identification and efficiency in developing devices.
Spin-torque oscillators have been applied to neuromorphic computing experimentally and theoretically, but the optimization of their physical parameters has usually been done by grid search.
In this paper, we propose a scheme to optimize the parameters of the dynamics of macrospin-type spin-torque oscillators using the gradient descent method with automatic differentiation.
First, we prepared numerically created dynamics as teacher data and successfully tuned the parameters to reproduce the dynamics.
This can be applied to obtain the correspondence between the simulation and experiment of the spin-torque oscillators.
Next, we successfully solved the image recognition task with high accuracy by connecting the coupled system of spin-torque oscillators to the input and output layers and training all of them through gradient descent.
This approach allowed us to estimate how to control the experimental setup and design the physical systems so that the task could be solved with a high accuracy using spin-torque oscillators.
\end{abstract}

\maketitle




The optimization of physical parameters has attracted attention from various perspectives, such as optimizing the physical parameters of simulations corresponding to experiments or maximizing the effect of a particular physical phenomenon, and the demand for such optimization has recently increased to identify suitable parameter regions for neuromorphic computing.
Physical reservoir computing \cite{tanaka2019recent,nakajima2020physical,nakajima2021physical} is a typical neuromorphic computing, where time series processing has been realized using various physical systems including photonic, \cite{fiers2013nanophotonic,van2017advances,brunner2019photonic,pauwels2019distributed,harkhoe2019task,nakajima2021scalable,shen2023deep} quantum, \cite{fujii2017harnessing,nakajima2019osaka,ghosh2021quantum,tran2021learning,fujii2021quantum,tran2023quantum,kubota2023temporal,kobayashi2024extending} and spintronic devices. \cite{grollier2020neuromorphic,grollier2016spintronic,torrejon2017neuromorphic,tsunegi2018evaluation,furuta2018macromagnetic,akashi2022coupled,tsunegi2023information,yamaguchi2023computational,bourianoff2018potential,raab2022brownian,nakane2018reservoir,watt2020reservoir,iihama2024universal}
Image recognition using spintronic devices has also been proposed recently. \cite{leroux2022convolutional,liu2024exploiting}




In general, models of physical systems used in neuromorphic computing are complicated and are characterized by many parameters.
For example, reservoir computing requires nonlinearities in the dynamics of the reservoir when the linear readout is used because a nonlinear transformation of the input time-series needs to be realized for nonlinear tasks.
Since the computational performance of such computing depends intricately on the physical parameters, it is difficult to specify the parameter values to increase computational performance.
It has commonly been shown that usual reservoir computing requires an echo state property, \cite{jaeger2001echo,yildiz2012re} where the reservoir state is expressed as a function of the previous input series only (in other words, the reservoir should forget the information of initial states).


Prior studies have investigated the physical parameter dependence of computational power by measuring it exhaustively throughout the entire parameter space.
For example, the dependence of the computational power of reservoir computing on physical parameters of spin-torque oscillators (STOs) has been investigated \cite{akashi2020input,imai2022input}.
In another example, the task-adaptive approach has been conducted to determine which tasks are suited to the ordered phases of magnetic materials. \cite{lee2024task}

In this paper, we propose the optimization of parameters in the simulation of a given physical system by using the gradient descent method with automatic differentiation.
The adjoint method with automatic differentiation has been investigated to make it possible to incorporate differential equations into the network and optimize the parameters that characterize those differential equations for a given objective (see supplementary material). \cite{chen2018neural,hughes2018adjoint,molesky2018inverse,nakajima2021neural,inui2024inverse,parsa2024gradient}
This method will allow us to determine the optimized parameter values of the simulation for the objective, and to suggest the experimental setup needed for that objective.
Before doing the ``optimization for tasks,'' we should do ``system identification.''
In this step, we match the dynamics of the simulation to the experimental data and thus we can obtain a correspondence between changes in the experimental setup and changes in the parameter values of the simulation.
Therefore, we will be able to apply simulation optimizations to experimental optimizations.
While the previous study \cite{chen2022forecasting} has used neural networks to emulate spintronics dynamics, here the spintronics dynamics itself is considered as the working network. 
Also, linear optimization to find the topology of spin texture that maximizes the magnetic field has also been done,\cite{bruckner2023magnum} but in this paper, a macrospin model is used to perform system identification and machine learning tasks, demonstrating advantages of our scheme in practical applications.

\begin{figure*}
\includegraphics[width=\textwidth]{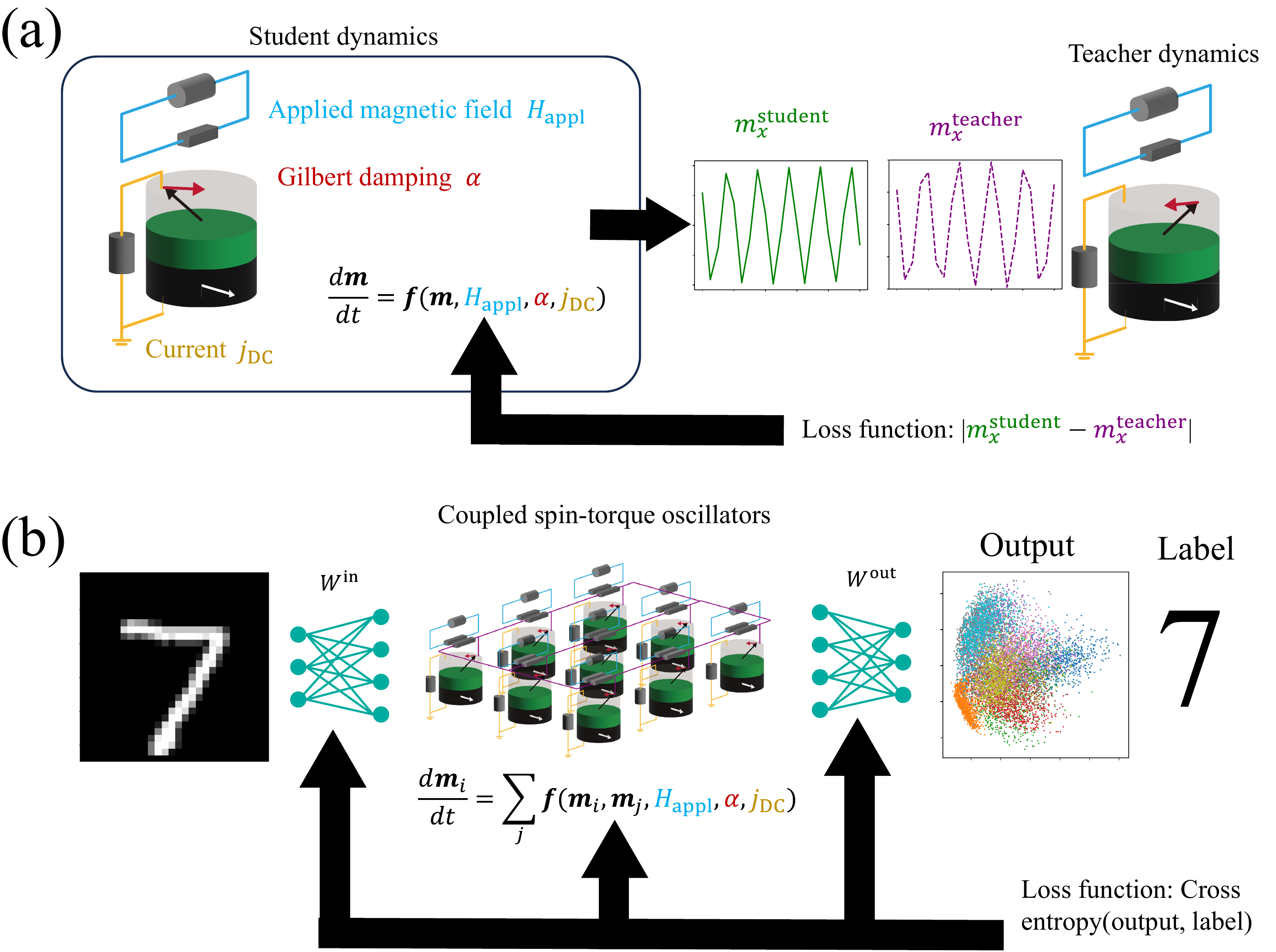}
\caption{\label{fig:fig1}Schematic illustration of (a) system identification of STOs and (b) optimization of the physical parameters of coupled STOs by using the gradient descent method with automatic differentiation. 
In the system identification, we predict the parameter value of the applied magnetic field, current, and Gilbert damping. 
In the optimization, we optimized the applied magnetic field, current, and Gilbert damping simultaneously, and/or readin and/or readout weight for the MNIST task.}
\end{figure*}

We apply our proposed method to the macrospin-type STOs that are described by the Landau-Lifshitz-Gilbert (LLG) equation.
The dynamics of the STOs are essentially a phenomenon described by a quantum many-body system, and it is difficult to derive a differential equation representing the dynamics rigorously.
In other words, the LLG equation is not a microscopic equation but a phenomenological equation in which it is difficult to get theoretical help in determining the parameter values.
Therefore, ``system identification'' is crucial in this case.
Figure \ref{fig:fig1}(a) schematically summarizes the concept of ``system identification'' in our study.
Then, as an example of ``optimization for tasks'', we solve the MNIST (Mixed National Institute of Standards and Technology database) task, which is a hand digit classification task, by optimizing the physical parameters of the coupled macrospin-type STOs with the input and output layer [Fig. \ref{fig:fig1}(b)].\cite{liu2024exploiting}

The model of the single macrospin-type STO is described as follows:
\begin{align}
    \frac{d{\bm m}}{dt} 
    = &-\gamma H_{\rm appl} {\bm m} \times {\bm e}_z 
    - \gamma (H_{\rm K} - 4\pi M) m_z {\bm m} \times {\bm e}_z 
    \nonumber\noindent\\
    &- \gamma H_s(t) {\bm m} \times ({\bm p} \times {\bm m}) 
    + \alpha {\bm m} \times \frac{d{\bm m}}{dt},
\end{align}
where ${\bm m}$ is the magnetization and
\begin{align}
    H_s(t) = \frac{\hbar \eta j_{\rm DC}}{2e(1+\lambda {\bm m}\cdot{\bm p})M\pi R^2 d},
\end{align}
$\hbar$ is the Dirac constant and $e$ is the elementary charge, ${\bm p}$ is the magnetization in the fixed layer with $x$ direction, and the parameter is summarized in Tab. \ref{tab:table1}.

We have three kinds of parameters: design parameters, internal parameters, and external parameters.
The design parameter is a parameter known to the experimenter at the time the substance is made: the radius and thickness of the STO.
The external parameter is a parameter that the experimenter manipulates from outside the substance: the applied magnetic field and current.
The internal parameter is a parameter determined by information inside the material, such as electrons and lattice: the gyromagnetic ratio, saturation magnetization, interfacial magnetic anisotropy field, spin polarization, spin-transfer torque asymmetry, spin polarization, and Gilbert damping.
The design parameter is a parameter whose value does not need to be estimated, while the internal parameter is a parameter whose value needs to be estimated.
The external parameter value also needs to be estimated; it is difficult to determine in detail how much current flows through the STO when the voltage is applied and how much magnetic field is induced by the current.

Another classification of these parameters is in which terms they contribute to the differential equation.
The LLG equation describes the precession of magnetization and is constructed by four terms.
Some parameters, such as the applied magnetic field appear in only one term, while others, such as the gyromagnetic ratio, appear in multiple terms.
We remark that some variables are equivalent in system identification. 
For example, because the current $j_{\rm DC}$ and the spin polarization $\eta$ contribute only to their multiplication $\eta j_{\rm DC}$ in the third term, predicting the value of the current is equivalent to predicting that of the spin polarization.
\begin{table}
\caption{\label{tab:table1}Summary of parameters of the LLG equation.}
\begin{ruledtabular}
\begin{tabular}{ccccc}
Symbol&Name&Category&Term\\
\hline
$\gamma$ & Gyromagnetic ratio & Internal & 1, 2, 3 & \\
$H_{\rm appl}$ & Applied magnetic field & External & 1 \\
$M$ & Saturation magnetization & Internal & 2, 3 & \\
$H_{\rm K}$ & Interfacial magnetic anisotropy field & Internal & 2 \\
$d$ & Thickness & Design & 3 & \\
$R$ & Radius & Design & 3 \\
$\eta$ & Spin polarization & Internal & 3 \\
$\lambda$ & Spin-transfer torque asymmetry & Internal & 3 \\
$j_{\rm DC}$ & Current & External & 3 \\
$\alpha$ & Gilbert damping & Internal & 4
\end{tabular}
\end{ruledtabular}
\end{table}
In this paper, we focus on the three parameters for the system identification: applied magnetic field $H_{\rm appl}$,  the direct current $j_{\rm DC}$, and the Gilbert damping ${\alpha}$.
In addition, we add the stochastic magnetic field ${\bm h}$ caused by thermal noise to the model where the correlation is defined as follows:
\begin{align}
\langle h_l (t) h_k (t') \rangle = \frac{2\alpha k_B T}{\gamma MV} \delta_{kl} \delta(t-t').
\end{align}
We set $\gamma=1.764\times10^7{\rm rad}/({\rm Oe}\cdot {\rm s})$, $M=1448.3 {\rm emu}/{\rm c.c.}$, $H_K=1.8616\times10^{4} {\rm Oe}$, $d=2.0\times10^{-7} {\rm cm}$, $R=60.0\times10^{-7} {\rm cm}$, $\eta=0.537$, $\lambda=0.288$, and $V$ is the volume.
In addition, we solve the LLG equation by the fourth-order Runge–Kutta method with a step width of $5.0\times10^{-13}$s.

Here, we prepare the teacher dynamics by using the simulation instead of the experiment.
We remark that the physical quantity actually measured in this case is the voltage, not the dynamics of magnetization. 
However, the voltage generally has a large proportional component of the $x$-component of the magnetization.
The loss function is given by the average of difference between the $x$-component of the teacher dynamics $m_x^{i,{\rm teacher}}$ and student dynamics $m_x^{i,{\rm student}}$ with respect to time and trials.
\begin{align}
    \mathcal{L} = \frac{1}{N_{\rm trial}T} \sum_{i=1}^{N_{\rm trial}} \sum_{t=0}^{T-1} |m_x^{i,{\rm teacher}}(t_{\rm sampling}t) - m_x^{i,{\rm student}}(t_{\rm sampling}t)|,
\end{align}
where we set the the sampling width $t_{\rm sampling}$ to $5.0\times10^{-11}$ s, the number of trials $N_{\rm trial}$ to 100, and the number of time steps $T$ to 10.
In the training phase, we match the initial student dynamics to the initial teacher dynamics and predict the parameter value of the teacher dynamics by the gradient descent method with the automatic differentiation.
We set the parameter range of the applied magnetic field to $1500 \ {\rm Oe} \le H_{\rm appl} \le 2000 \ {\rm Oe}$, the current to $2.5 \ {\rm mA} \le j_{\rm DC} \le 7.5 \ {\rm mA}$, and the Gilbert damping to $0.001 \le \alpha \le 0.01$.
We set washout duration for teacher (student) dynamics to 100 (0.5) ns.

\begin{figure*}
\includegraphics[width=\textwidth]{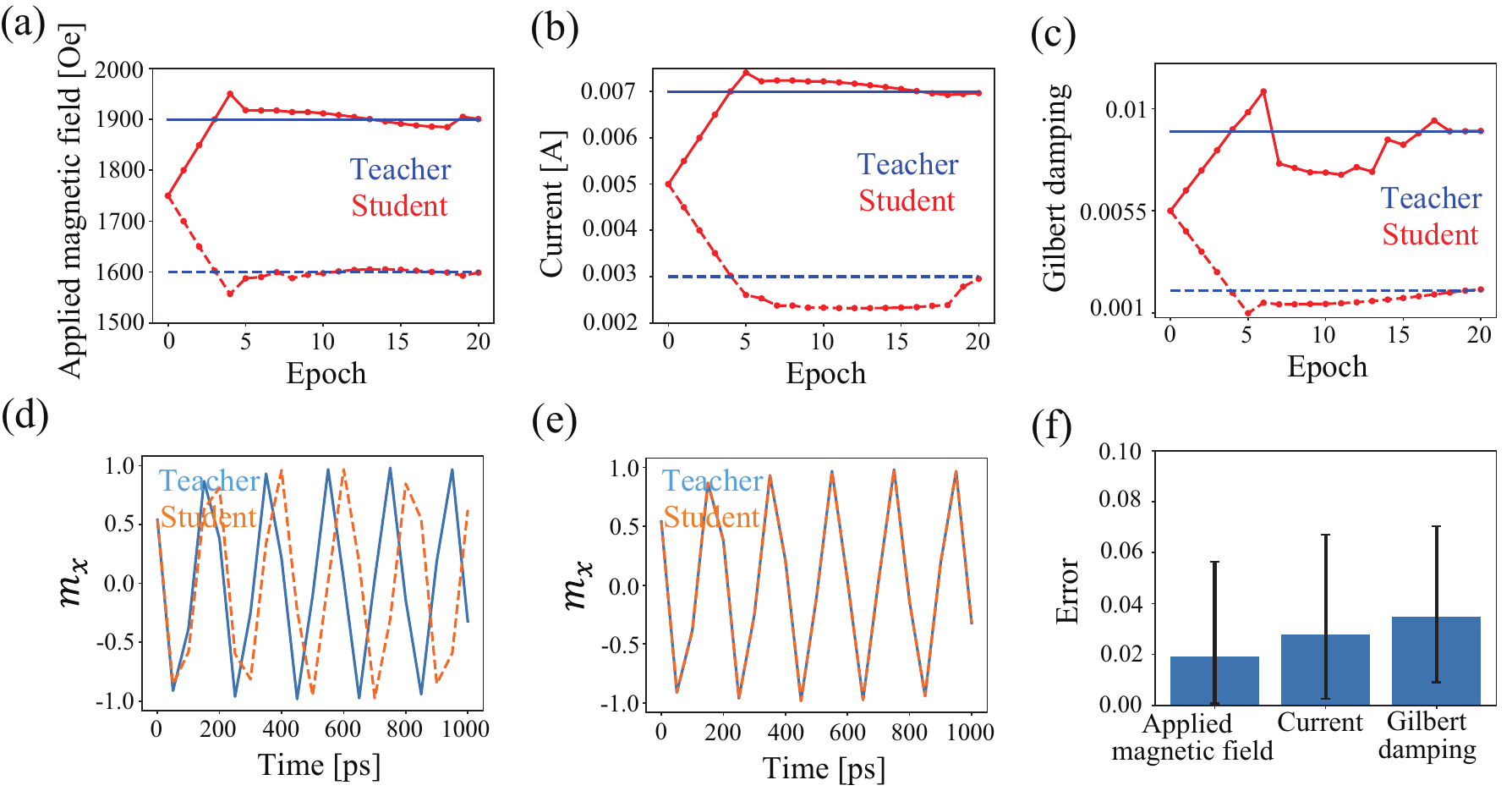}
\caption{\label{fig:fig2}The result of system identification for the STO. Epoch dependence of (a) applied magnetic field, (b) current, and (c) Gilbert damping for two cases (solid and dashed line), where the blue (red) line shows the Gilbert damping for the teacher (student) dynamics. The teacher (orange) and student (blue) dynamics (d) before and (e) after the training in the optimization of the Gilbert damping. (f) Averaged error of 10 cases in 20 epochs for the applied magnetic field, current, and Gilbert damping.}
\end{figure*}

Figure \ref{fig:fig2} shows the results of system identification where the learning rate is basically set by 1/10 of the parameter range and the learning rate is set to halve in cases of rising losses during the learning.
In Fig. \ref{fig:fig2}(a)–(c), we plot the successful two cases of system identification for each parameter optimization of the applied magnetic field, current, Gilbert damping, respectively.
We initially set the unknown parameter value of 100 trials to the middle of the parameter range.
We then plot the parameter values with the lowest loss among the 100 trials. 
The model was successfully estimated at 20 epochs when the correct parameter values were moderately far from the initial parameter values.
Figure \ref{fig:fig2}(d) [(e)] shows the two dynamics in the case of system identification for the Gilbert damping before (after) the learning and shows that the two dynamics become close by the learning.
Figure \ref{fig:fig2}(f) shows the averaged error of 10 cases in 20 epochs, where the error is defined by the difference between the correct parameter value and the optimal parameter value divided by the width of the parameter range.
The error depends on the difference between the correct parameter value and the initial parameter value and on the learning rate (see supplementary material).

\begin{figure*}
\includegraphics[width=\textwidth]{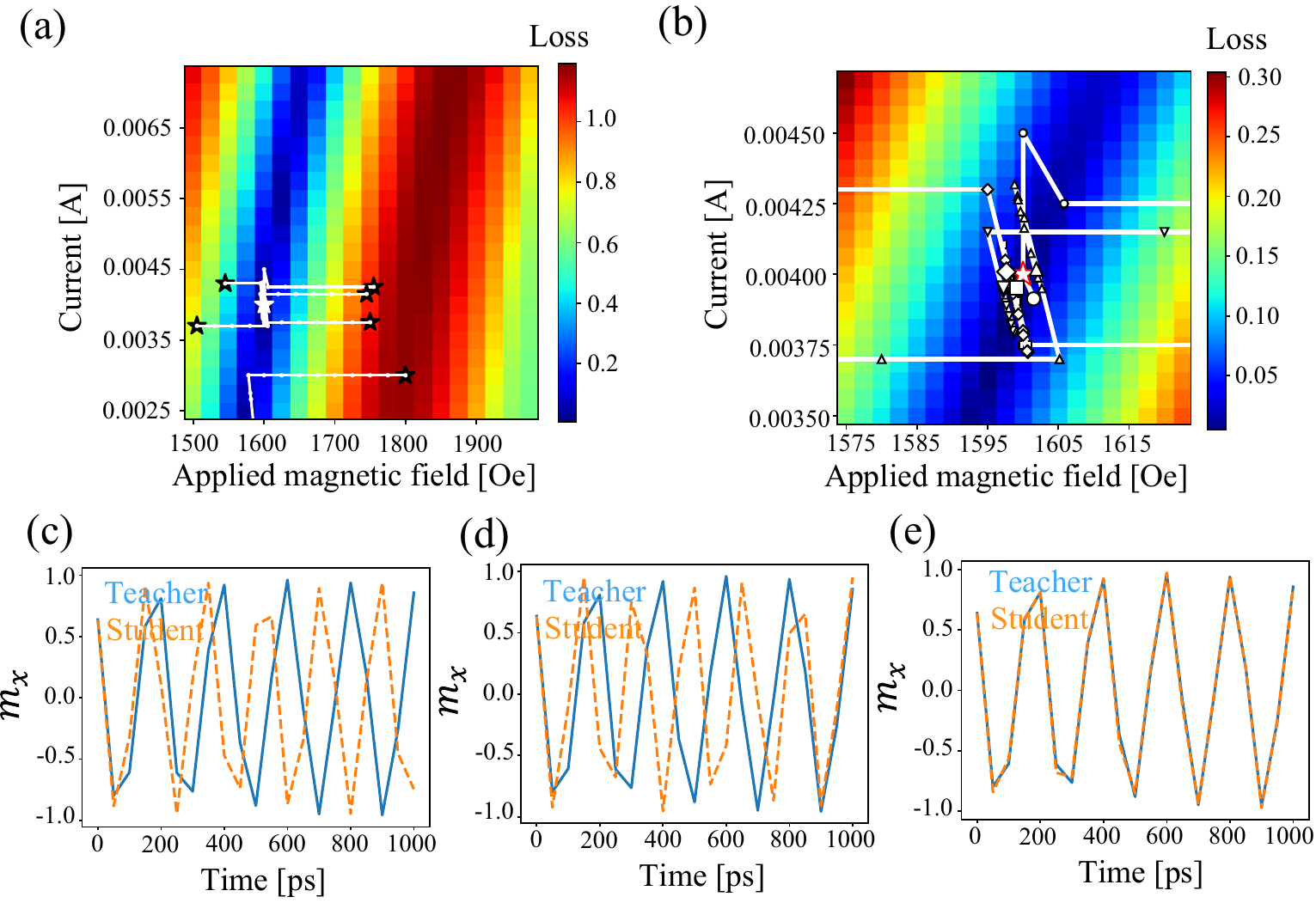}
\caption{(a) The dependence of loss function on current and applied magnetic field where the white lines show the change of the parameter values during the training and the black star points show (white star point shows) parameter values before (after) the training.
(b) The enlarged view of (a) where the final parameter values are represented by the large points.
Teacher (blue) and initial student (orange) dynamics  where the current and applied magnetic field are (c) 0.004 A, 1825 Oe, (d) 0.004 A, 1975 Oe, (e) 0.0055 A, 1625 Oe.
\label{fig:fig3}}
\end{figure*}

Figure \ref{fig:fig3} shows the case of two-parameters prediction for current and applied magnetic field where the Gilbert damping constant is set to 0.0055.
Figures \ref{fig:fig3}(a) and (b) show the dependence of the loss on the current and applied field.
When the initial value of the current is close to the correct value, the parameters were successfully estimated by the following two steps: (1) tuning only applied magnetic field with the learning rate of 25 until the minimum loss with respect to the batch reaches to 0.15, (2) tuning the applied magnetic field and current with the learning ratio of 2.5 and 0.00025.
However, if the initial current value is far from the correct value, learning in a similar manner will lead toward a different local solution.
This can be overcome by introducing a scheduling scheme of the learning rates, which is included in our future work.
Figure \ref{fig:fig3}(c)-(e) shows the teacher and student dynamics where the loss function is large, moderate, and small.
Although systematic analysis is difficult because of the transient dynamics used in the student dynamics \cite{taniguchi2017relaxation}, it can be inferred that many minima are created by the balance between the current directing the magnetization in-plane and the magnetic field directed perpendicularly.

Then we solve the MNIST tasks by connecting the coupled macrospin-type STOs to the input layer $W^{\rm in}$ and output layer $W^{\rm out}$ [Fig. 1(b)] by the following formula:
\begin{align}
  {\bm x} (0) &= W^{\rm in} {\bm u} (n),
\\
  {\bm x} (t) &= {\bm x} (0) + \int_0^t {\bm x} (s) ds ,
\\
  {\bm y} (n) &= W^{\rm out} [{\bm x} (T)^{\top} ; 1]^{\top},
\end{align}
where ${\bm x}$ is a given dynamical system (in our case, the coupled macrospin-type STOs), ${\bm u}$ is the input data, and ${\bm y}$ is the output.
The coupling magnetic field for $i$th STO is defined by 
\begin{align}
    {\bm H}_{i} = A_{\rm cp} \sum_{k=1}^{N} w_{i,k}^{\rm cp} m_{k,x} {\bm e}_x,
\end{align}
where $N=200$ and $w_{i,k}$ is an element of the matrix normalized so that the spectral radius is 1 \cite{akashi2022coupled}.
Therefore, $A_{\rm cp}$ determines the spectral radius of the coupling magnitude (CM).
Here, we optimize the applied magnetic field, current, and Gilbert damping, and also study the dependence of the accuracy on the CM.
We also set the different applied magnetic fields for the different STOs and the same current and Gilbert damping for all the STOs.
Furthermore, we set the step width to 50 ps and the iteration time to 500 ps, the learning rate for the applied magnetic field to 1.0, and the current and Gilbert damping to 1.0$\times10^{-6}$, and the input and output layers to 1.0$\times10^{-3}$.
Also, we used $x$, $y$, and $z$ component of the dynamics of the macrospin-type STO as the output of the physical layer.
In addition, we used cross entropy as a loss function and decayed the learning ratio by the factor of 0.85.

\begin{figure*}
\includegraphics[width=\textwidth]{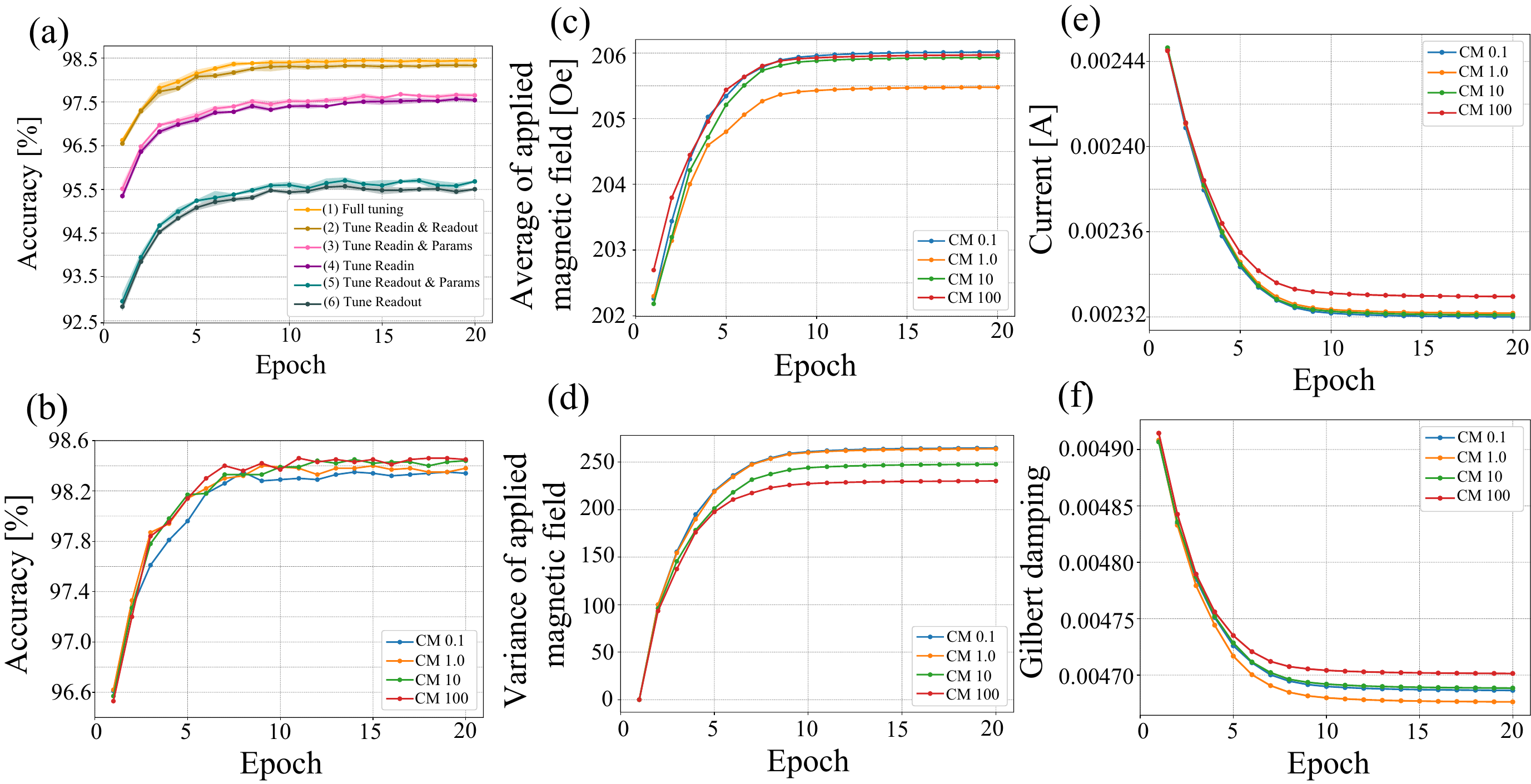}
\caption{\label{fig:fig4} (a) Accuracy with (1) the full tuning, (2) tuning readin and readout weight, (3) tuning readin and physical parameters, (4) tuning readin weight, (5) tuning readout and physical parameters, and (6) tuning readout weight.
(b) Epoch dependence of the accuracy when the CM equals to 0.1, 1.0, 10, and 100.
The epoch dependence of (c) average and (d) variance of applied magnetic field, (e) current, and (f) Gilbert damping for the MNIST task when the CM equals to 0.1, 10.0, 10, and 100.}
\end{figure*}

In Fig. \ref{fig:fig4}(a), we show the averaged accuracy with respect to the CM (CM = 0.1, 1.0, 1.0, and 100).
We studied six cases: (1) tuning readin weight, physical parameters, and readout weight (full tuning), (2) tuning readin and readout weights, (3) tuning readin weight and physical parameters, (4) tuning readin weight, (5) tuning readout weight and physical parameters, and (6) tuning readout weight.
Cases (1) and (2) show the increase in accuracy by learning the physical parameters as well, rather than learning only the readin and readout weight. 
The readout/readin weight is not trained in the case (3)/(5) and (4)/(6), and the accuracy is increased by learning the physical parameters as well, rather than learning only the readin/readout weight.
Furthermore, Fig. \ref{fig:fig4}(a) shows that tuning readin has a greater impact on accuracy than tuning readout.
Figure \ref{fig:fig4}(b) shows the dependence of the accuracy on the CM.
The accuracy is enhanced when the CM is large because the coupling makes the dynamics of each STO independent and increases the convertibility due to the bifurcation in the limit cycle and chaotic attractor.\cite{liu2024exploiting}
The dynamics realized by the coupled macrospin-type STOs are shown in supplementary material.
We remark that fixing the input or output layer means that the input or output layer can be physicalized, which leads to the replacement of the conventional computer by other physical systems.
Figure \ref{fig:fig4}(c)-(f) shows the convergence of the physical parameters, where all the parameters are tuned.
One can see that the average of the applied magnetic field, current, and Gilbert damping converged to almost the same value for each.
Figure 4(d) shows the variance of the applied magnetic fields for the full tuning.
It can be seen that there was no variance before the training, but variance was created by the training.
The comparison of the convergence of the physical parameters for the full tuning and the tuning other than the readout is shown in supplementary material.
Table \ref{tab:table2} shows the summary of accuracy of MNIST task for all the configurations of the training shown in Fig. \ref{fig:fig3}(a) [(1)-(6)] where the CM is 0.1, 1.0, 10.0, 100.0.
Whatever the value of CM, the effectiveness of tuning of the STOs was demonstrated.

\begin{table}
\caption{\label{tab:table2}Summary of accuracy for a single trial of MNIST task where (1)-(6) correspond to those in Fig. \ref{fig:fig4}. The upper right corner has the largest accuracy, which is comparable to the accuracy (98.77 at highest) by a convolutional neural network, having the same computational nodes (600 nodes) in a middle layer.\cite{liu2024exploiting}}
\begin{ruledtabular}
\begin{tabular}{c|cccc}
 & CM=0.1 & CM = 1.0 & CM = 10.0 & CM = 100.0 
\\ \hline 
(1) & 98.380 & 98.480 & 98.420 & 98.490 
\\ 
(2) & 98.379 & 98.320 & 98.350 & 98.270 
\\ 
(3) & 97.700 & 97.600 & 97.660 & 97.630 
\\ 
(4) & 97.470 & 97.570 & 97.570 & 97.540 
\\ 
(5) & 95.700 & 95.700 & 95.660 & 95.670 
\\ 
(6) & 95.490 & 95.500 & 95.490 & 95.540  
\end{tabular}
\end{ruledtabular}
\end{table}

We showed system identification for the macrospin-type STO by the gradient descent method with automatic differentiation.
For system identification to be perfect, the model must be perfectly consistent with the experiment.
However, such consistency is almost impossible in reality.
The STO treated in this paper is a quantum many-body system, which limits its modeling due to its high complexity.
We focused on the macrospin model as a first step, but a more rigorous model, such as the micromagnetic model, can be applied to system identification.\cite{bruckner2023magnum}
Further theoretical development is needed to make the model more rigorous.
However, system identification does not necessarily aim to perfectly align the model with the experiment.
For example, if the goal is to find physical parameters that increase the performance of computing, it is sufficient if the model and experiment are identical in terms of computing performance.
Thus, the inconsistency between the model and experiment would not necessarily be a problem for calculation purposes.
In addition, we showed that the optimization of physical parameters of the coupled macrospin-type STOs for the MNIST task can be done. 
Our method can also be applied to time-series processing by reservoir computing with the STO.

\section*{Supplementary infromation}

\subsection*{Optimmizing system parameters using adjoint method}
We show details of the optimization method.
As for the system identification, physical parameters such as the applied magnetic field, current, and Gilbert damping are optimized.
The LLG equation including a parameter $P$ can be represented as follows:
\begin{align}
    \frac{d{\bm m}}{dt} = {\bm f} ({\bm m}, P),
\end{align}
where ${\bm f}$ can be expressed by the following form excluding the time derivative:
\begin{align}
    {\bm f} = - \frac{\gamma}{1 + \alpha^2} {\bm m} \times {\bm b} - \frac{\alpha\gamma}{1+\alpha^2} {\bm m} \times ({\bm m} \times {\bm b} ),
\end{align}
where
\begin{align}
    {\bm b} = {\bm H} + \frac{\gamma \hbar \eta j}{2eMd} {\bm p} \times {\bm m}.
\end{align}
Then, we define the adjoint ${\bm a}$ as follows:
\begin{align}
    {\bm a} = \frac{\partial \mathcal{L}}{\partial {\bm m}},
\end{align}
with the loss function $\mathcal{L}$ defined in the main text. 
The time evolution of the ajoint is given by
\begin{align}
    \frac{d{\bm a}}{dt} = - {\bm a}^T \frac{\partial {\bm f}}{\partial {\bm m}} .
\end{align}
Similarly, one can obtain the following relation.
\begin{align}
    \frac{d}{dt} \frac{\partial \mathcal{L}}{\partial P} = - {\bm a}^T \frac{\partial {\bm f}}{\partial P}.
\end{align}
To obtain the loss function, we first solve the LLG equation forwardly.
Next, by solving the above equations in the opposite direction, the back propagation can be achieved.
We optimized the parameter $P$ by Adam:
\begin{align}
    P &\leftarrow P - \frac{\eta \hat{m}}{\sqrt{\hat{v}+\epsilon}},
    \\
  \hat{m} &= \frac{m}{1-\beta_1},
\\
  \hat{v} &= \frac{v}{1-\beta_2},
\\
  m &\leftarrow \beta_1 m + (1-\beta_1)\frac{\partial \mathcal{L}}{\partial P},
\\
  v &\leftarrow \beta_2 v + (1-\beta_2) \left( \frac{\partial \mathcal{L}}{\partial P} \right)^2,
\end{align}
where we set $\beta_1=0.9$, $\beta_2=0.999$, $\epsilon=1.0\times10^{-8}$ and we set the learning ratio $\eta$ depending on the case.
As for the one-parameter prediction, we set the learning ratio for a parameter $P$ to a parameter range of $P$ multiplied by 0.1.
As for the two-parameters prediction of the applied magnetic field and current, we first optimized the applied magnetic field with the learning ratio of 25 until the minimum loss with respect to the batch reaches to 0.15.
Then, we optimized the applied field and current with the learning ratio of 2.5 and 0.00025.

To solve the MNIST task, we connected the coupled STOs to the input layer and output layer where the norm of magnetization is always set to 1.
In this task, we optimized the readin weight ($W^{\rm in}$), the STO parameters (applied magnetic field, current, and Gilbert damping), and the readout weight ($W^{\rm out}$).
We used Adam repeatedly regarding batch to optimize the parameters where $\beta_1=0.9$, $\beta_2=0.999$, $\epsilon=1.0\times10^{-8}$.
We set the learning ratio $\eta$ for input and out layer to 1.0$\times10^{-3}$, current and Gilbert damping to 1.0$\times10^{-6}$.
We also decayed the learning ratio by the ratio of 0.85.

\subsection*{Effects of thermal noise on the system identification}
\begin{figure*}
\includegraphics[width=\textwidth]{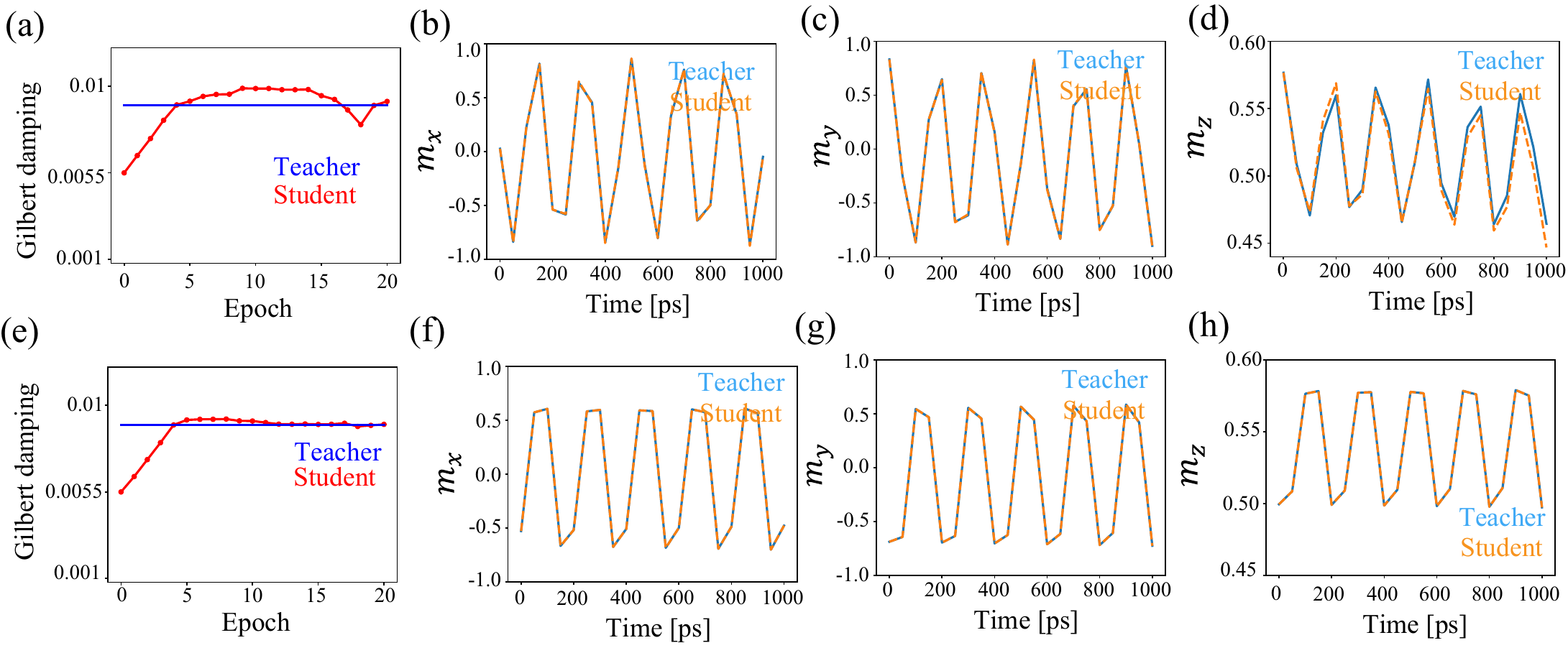}
\caption{\label{fig:figS1}(a) Epoch dependence of Gilbert damping with the thermal noise where the blue (red) line shows the Gilbert damping for the teacher (student) dynamics. (b)/(c)/(d) The $x$/$y$/$z$-component of the teacher (blue) and student (orange) dynamics after the training with the thermal noise. (e) Epoch dependence of Gilbert damping without the thermal noise where the blue (red) line shows the Gilbert damping for the teacher (student) dynamics. (f)/(g)/(h) The $x$/$y$/$z$-component of the teacher (blue) and student (orange) dynamics after the training without the thermal noise.}
\end{figure*}

In the main text, we show the result of the system identification with the thermal noise.
Here, we compare the system identification for the Gilbert damping with the thermal noise and without the thermal noise.
Figure \ref{fig:figS1}(a) [(e)] shows the epoch dependence of the Gilbert damping with (without) the thermal noise.
The physical parameters converge more smoothly in the absence of thermal noise than in the presence of thermal noise.
Figure \ref{fig:figS1}(b)–(d) [(f)–(h)] shows the dynamics after the training with (without) the thermal noise.
One can see that the periodicity of the dynamics is disturbed by thermal noise.
In the absence of thermal noise, the student and teacher dynamics match, but in the presence of thermal noise, the dynamics do not match due to differences in thermal noise realizations.
This difference is clearly evident in the $z$-component of the dynamics.
Despite such differences in dynamics due to thermal noise, system identification works well.

\subsection*{System identification by various learning rate and teacher dynamics}
\begin{figure*}
\includegraphics[width=\textwidth]{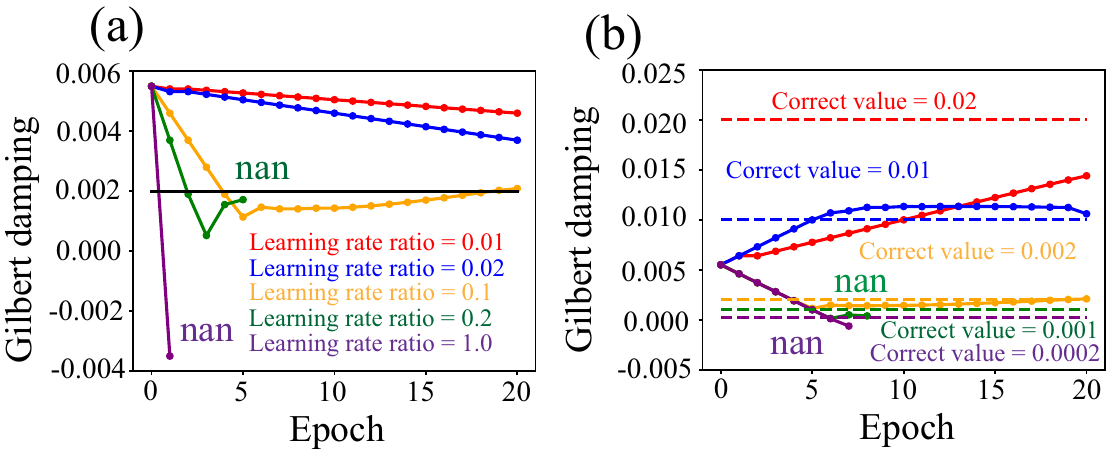}
\caption{\label{fig:figS2}(a) Epoch dependence of Gilbert damping for various learning rate ratios where ``nan'' shows that the training fails due to the physical unreality.
(b) Epoch dependence of Gilbert damping for various correct values.}
\end{figure*}

We show various case of system identification for the Gilbert damping, including the extreme cases.
Here, we define the learning rate ratio by the division of the learning rate by the range of parameter value being considered.
In Fig. \ref{fig:figS2}(a), we set the parameter range of the Gilbert damping to $0.001\le\alpha\le0.01$ and show the epoch dependence of the Gilbert damping for various learning ratios.
In Fig. \ref{fig:figS2}(b), we set the learning rate to 0.009 and show the epoch dependence of the Gilbert damping for various correct values.
Generally, if the learning rate is made too large and the Gilbert damping goes negative, ``nan'' will occur due to physical unreality.
See the green (learning rate ratio = 0.2) and purple line (learning rate ratio = 1.0) in Fig. \ref{fig:figS2}(a) and the green (correct value = 0.001) and purple lines (correct value = 0.0002) in Fig. \ref{fig:figS2}(b).
Conversely, if the learning rate is too small, it takes a long epoch for the physical parameters to converge to the correct values.
See the red (learning rate ratio = 0.01) and blue (learning rate ratio = 0.02) lines in Fig. \ref{fig:figS2}(a) and the red line (correct value = 0.020) in Fig. \ref{fig:figS2}(b).
We found that the training worked well in about 20 epochs, when the learning rate was about 1/10 of the given parameter range and obtained the results shown in the main text.

\section*{Convergence of physical parameter values for MNIST task}
\begin{figure*}
\includegraphics[width=\textwidth]{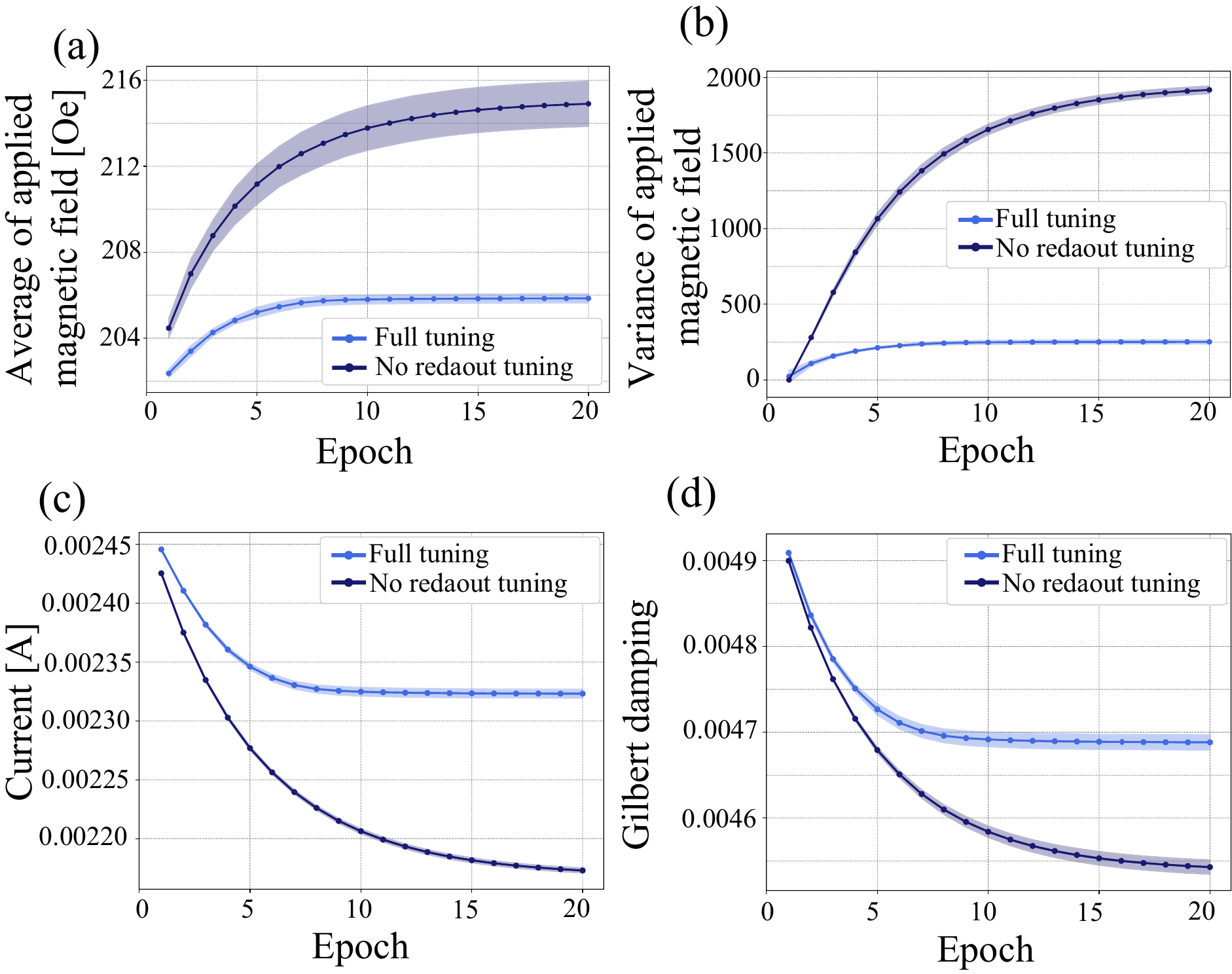}
\caption{\label{fig:figS3} The epoch dependence of (a) average and (b) variance of applied magnetic field, (c) current, and (d) Gilbert damping for the MNIST task by the full tuning (light blue) and tuning other than the readout layer (dark blue) where the spread around the line shows dependence on CM (CM = 0.1, 1.0, 10, 100).}
\end{figure*}

We compare the convergence of physical parameter values for the MNIST task by the full tuning and the tuning other than the readout layer. 
Figure \ref{fig:figS3}(a), (c), (d) shows the epoch dependence of the average of applied magnetic field, current, Gilbert damping by the full tuning and the tuning other than the readout layer.
It can be suspected that when read-out is frozen during training, the physical parameters have more impact to the system's accuracy and hence being tuning more effectively.
Figure \ref{fig:figS3}(b) shows the variance of the applied magnetic fields for the full tuning and tuning other than the readout layer.
In both cases, one can see the the variance is created by learning.

\subsection*{Dynamics of coupled spin-torque oscillators}
\begin{figure*}
\includegraphics[width=\textwidth]{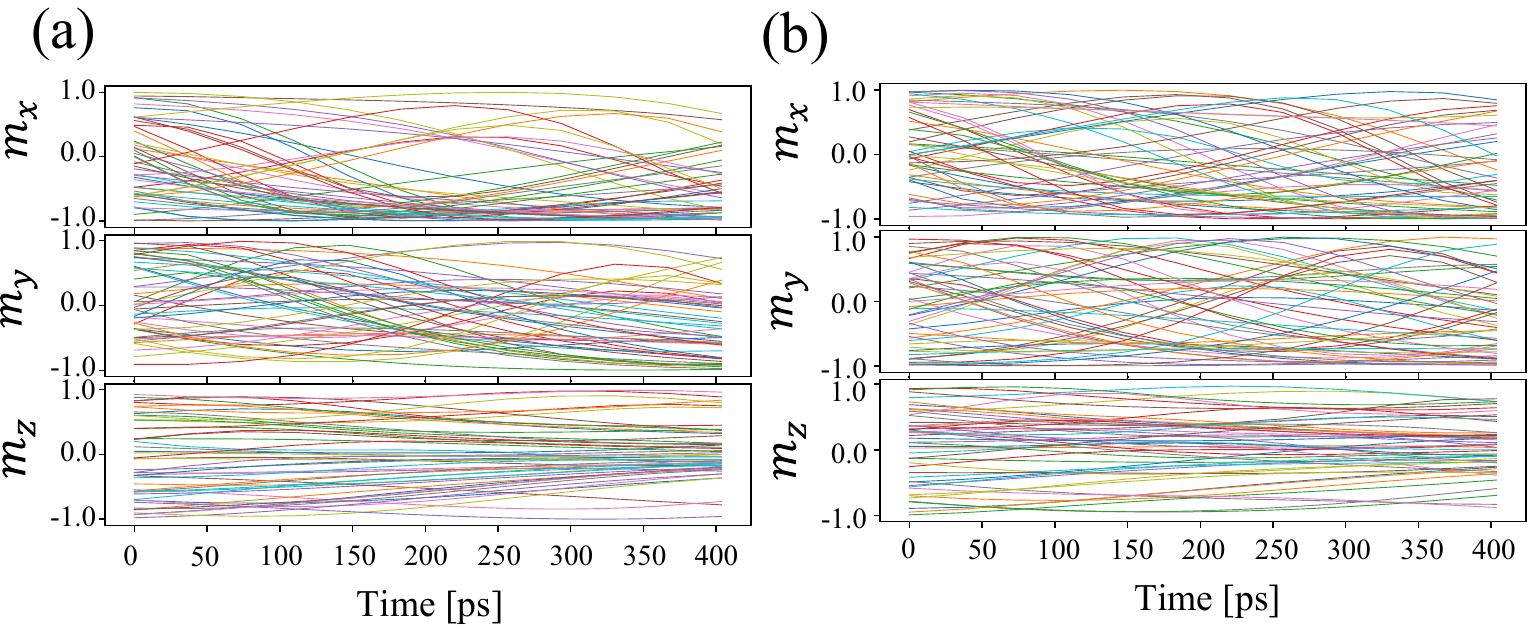}
\caption{\label{fig:figS4} (a)/(b)/(c) [(d)/(e)/(f)] The $x$/$y$/$z$-component of the dynamics of coupled macrospin-type spin-torque oscillators before (after) the training.}
\end{figure*}

The dynamics of coupled macrospin-type spin-torque oscillators are shown before (after) the training in Fig. \ref{fig:figS4}(a) [(b)].
We found that the overlap of dynamics among the spin-torque oscillators is reduced by the training.
This diversity of dynamics is important for solving the task because it increases the convertibility of the input data.

\nocite{*}
\bibliography{aipsamp}

\end{document}